\documentclass[twocolumn,showpacs,prd]{revtex4}
\usepackage{mathrsfs}
\usepackage{amsmath, txfonts}
\usepackage{graphicx,,booktabs,bm}
\usepackage{longtable,lscape}
\usepackage{overpic,multirow}
\usepackage{color}
\usepackage{amssymb}
\def\half{{\textstyle{1\over 2}}}
\def\thalf{{\textstyle{3\over 2}}}

\usepackage{indentfirst}
\begin{document}
\title{Study of the $B\bar{B}^*/D\bar{D}^*$ bound states in a Bethe-Salpeter approach}
\author{Jun He}\email{junhe@impcas.ac.cn}
\affiliation{
Theoretical Physics Division, Institute of Modern Physics, Chinese Academy of Sciences,Lanzhou~730000,China
}

\affiliation{
Research Center for Hadron and CSR Physics,
Lanzhou University and Institute of Modern Physics of CAS, Lanzhou 730000, China
}

\affiliation{
State Key Laboratory of Theoretical Physics, Institute of
Theoretical Physics, Chinese Academy of Sciences, Beijing  100190,China
}

\begin{abstract}

In this work the $B\bar{B}^*/D\bar{D}^*$ system is studied in the
Bethe-Salpeter approach with quasipotential approximation.  In our calculation both
direct and cross diagrams are included in the one-boson-exchange potential. The numerical
results indicate the existence of an isoscalar bound state $D\bar{D}^*$
with $J^{PC}=1^{++}$, which may be related to the $X(3872)$. In the
isovector sector, no bound state is produced from the
interactions of $D\bar{D}^*$ and $B\bar{B}^*$, which suggests the
molecular state explanations for $Z_b(10610)$ and $Z_c(3900)$ are
excluded.

\end{abstract}

\maketitle
\maketitle

\section{Introduction}

The deuteron is a loosely bound state of two nucleons.  It is natural to
expect other bound states composed of two hadrons, that is hadronic
molecular state~\cite{Tornqvist:1991ks}.  After the observation by the
Belle Collaboration \cite{Choi:2003ue} the $X(3872)$ was related to a
loosely bound state of $D\bar{D}^*$ immediately
\cite{Tornqvist:2004qy,Close:2003sg,AlFiky:2005jd} due to its mass near the
$D\bar{D}^*$ threshold. Recently, the Belle Collaboration announced
two charged bottomonium-like structures $Z_b(10610)$ and $Z_b(10650)$
near the $B\bar{B}^*$ and  $B^*\bar{B}^*$ thresholds~\cite{Belle:2011aa}.
The analysis of the angular distribution indicated both $Z_b(10610)$
and $Z_b(10650)$ favor $I^G(J^P)=1^+(1^+)$. A structure $Z_c(3900)$
close to the $D\bar{D}^*$ threshold was also observed by the BESIII
collaboration in the decay of $Y(4260)$, $Y(4260)\to\pi^+\pi^-
J/\psi$~\cite{Ablikim:2013mio}.

In Refs.~\cite{Sun:2011uh,Sun:2012zzd}, the $B\bar{B}^*/D\bar{D}^*$
system was studied with a nonrelativistic one-boson-exchange (OBE)
model by solving the Sch\"odinger equation. There exists a bound state
solution with quantum number $J^{PC}=1^{++}$ from  $B\bar{B}^*$
interaction while there exists no bound state solution from
$D\bar{D}^*$ interaction. The results also suggested the importance of
$\pi$ exchange~\cite{Sun:2012zzd}. It is easy to understand because
the binding energy is small for a hadronic molecular state so that
long range interaction should be more important than short range
interaction. In the nonrelatvistic OBE model, potential
$V({\bm r})$ is obtained with a direct Fourier transformation on ${\bm
q}$~\cite{Tornqvist:1991ks,Sun:2011uh}.
However, for hadronic molecular state, a system  composed of two
constituents with different masses and/or spins is often involved,
which is different from the deuteron where proton and neutron are
indistinguishable under isospin $SU(2)$ symmetry. Such difference may
lead to invalidness of the potential model in coordinate space,
which has been ignored always.

The molecular state is a loosely bound state of two hadrons, so the
Bethe-Salpeter equation (BSE) is an appropriate tool to deal with the
molecular state.  For example, BSE was used to study deuteron  and experimental data about deuteron and nucleon-nucleon
interaction were well reproduced~\cite{Gross:2010qm,Hummel:1989qn}. In
Ref. \cite{Ke:2012gm} the $B\bar{B}^*$ system has been studied in the BSE
approach with a quasipotential approximation. However, in their study
only direct diagram was included in the calculation. Hence, the most
important $\pi$ exchange as found in Ref.~\cite{Sun:2012zzd}, was not
included.  In Ref.~\cite{He:2011ed} the $Y(4274)$ and its three body
decay were studied in the BS equation approach with nonrelativistic
approximation \cite{He:2011ed}. And this method was successfully
applied to the $D^*_0(2400)N$ system. The $\Sigma_c(3250)$
reported by the BABAR collaboration recently can be explained as a
$D^*_0(2400)N$ molecular state \cite{He:2012zd}.  In this work, we
will develop a relativistic theoretical frame in BES approach to study
the $B\bar{B}^*/D\bar{D}^*$ system with $\pi$ exchange and search bound sate solution to understand the structures $Z_b(10610)$ and
$Z_c(3900)$.

This work is organized as follows. In next section we present
theoretical frame to study the $B\bar{B}^*/D\bar{D}^*$
system through solving the BSE. In Sec. III, the OBE potential is
derived with the help of the effective Lagrangian from the heavy quark
effective theory. The numerical results are given in Sec. IV. In
the last section, a brief summary is given.

\section{BSE with quasipotential approximation}
The BSE of the vertex function $\Gamma\rangle$ in general form is ~\cite{Adam:1997cx}
\begin{eqnarray}
	&&|\Gamma\rangle={\cal V} G~ |\Gamma\rangle,
\end{eqnarray}
where ${\cal V}$ and $G$ are the potential kernel and the propagator.

\subsection{BSE with definite quantum numbers}

\begin{figure*}[hbtp!]
\begin{center}
\includegraphics[bb=130 525 590 700,scale=0.9,clip]{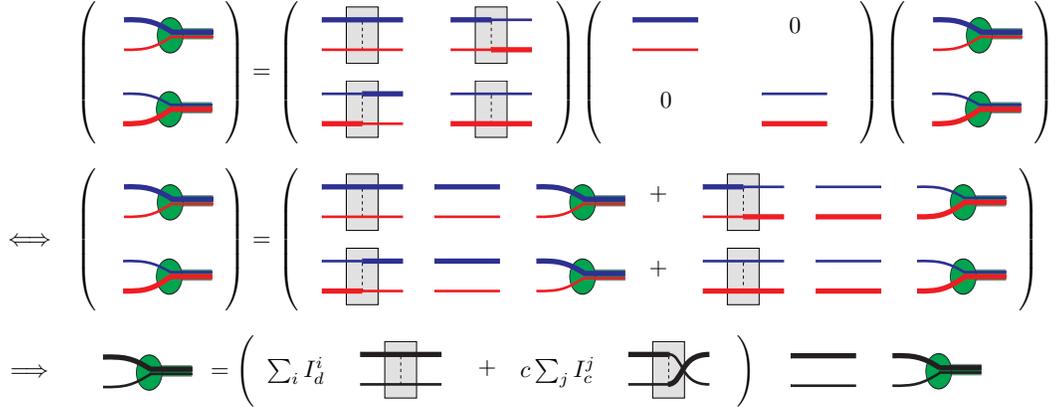}
\end{center}
\caption{(Color online) The BS equation for
	$|{Z_{P\bar{P}^*}^{(T)}}^+\rangle$. The thick and thin lines
	are for pseudoscalar and vector mesons, respectively. The red and
blue lines are for charged and neutral mesons, respectively. The black
lines are for the
diagram after isolating the flavor factors to $I^i$. In the last line the $SU(2)$ symmetry is applied.\label{Fig: BS}}
\end{figure*}

The $P\bar{P}^*$ systems  (here and hereafter we mark $B$ or $D$ as $P$) can be
categorized as the isovector ($T$) and isoscalar ($S$) states under SU(3) symmetry with
corresponding flavor wave functions,~\cite{Sun:2011uh,Sun:2012zzd}
\begin{eqnarray}
&&\left\{\begin{array}{l}
|{Z_{P\bar{P}^*}^{(T)}}^+\rangle=\frac{1}{\sqrt{2}}\big(|P^{*+}\bar{P}^0\rangle+cP^+\bar{P}^{*0}\big),\\
|{Z_{P\bar{P}^*}^{(T)}}^-\rangle=\frac{1}{\sqrt{2}}\big(|P^{*-}\bar{P}^0\rangle+cP^-\bar{P}^{*0}\big),\\
|{Z_{P\bar{P}^*}^{(T)}}^0\rangle=\frac{1}{2}\Big[\big(|P^{*+}P^-\rangle-P^{*0}\bar{P}^0\big)
+c\big(P^+P^{*-}-P^0\bar{P}^{*0}\big)\Big],\end{array}\right. \label{e1}\\
&&|{Z_{P\bar{P}^*}^{(S)}}^0\rangle=\frac{1}{2}\Big[\big(|P^{*+}P^-\rangle+P^{*0}\bar{P}^0\big)
+c\big(P^+P^{*-}+P^0\bar{P}^{*0}\big)\Big],\label{e2}
\end{eqnarray}
where $c=\pm$ corresponds to $C$-parity $C=\mp$ respectively.

Now we introduce the BSE with definite quantum numbers especially isospin $I$ and charge parity $C$.
Here we take the positive charged system
$|{Z_{P\bar{P}^*}^{(T)}}^+\rangle$ as example to explain how to obtain the BSE for a system with definite quantum numbers.   First, the coupled channel
BSE for the two-component of
the positive charged system,
\begin{eqnarray}
|{Z_{P\bar{P}^*}^{(T)}}^+\rangle=\frac{1}{\sqrt{2}}
\big(~\raisebox{-6pt}{\includegraphics[ bb=100 530 500
750,scale=0.11,clip]{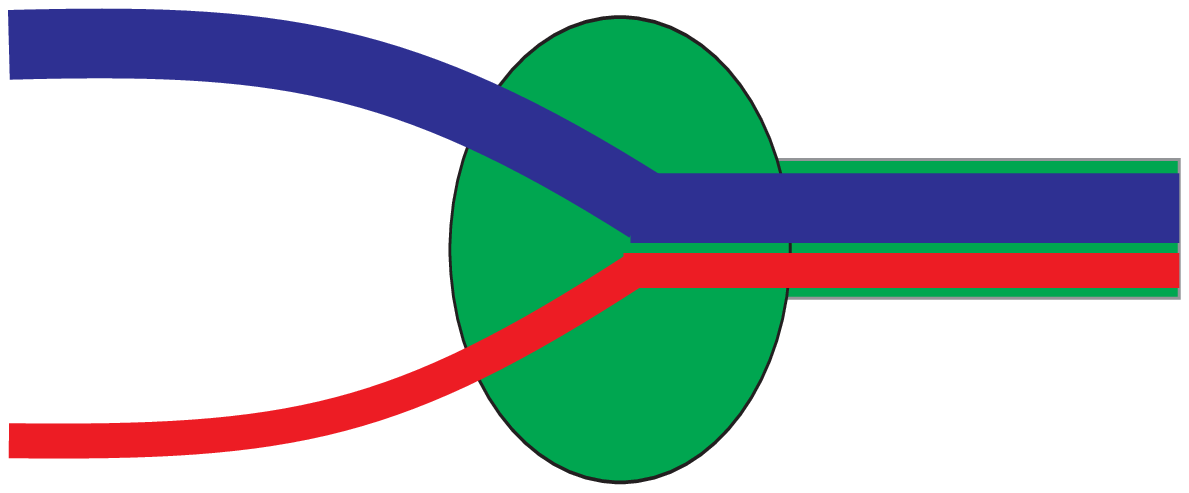}}+c\raisebox{-6pt}{\includegraphics[
bb=100 530 500
750,scale=0.11,clip]{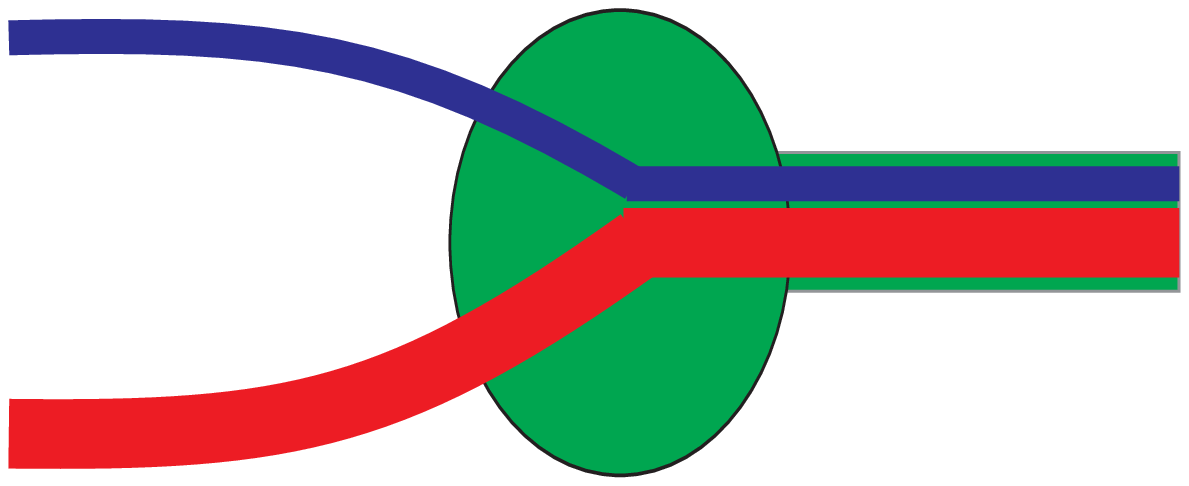}}~\big).\label{Eq:positive charged
sytstem}
\end{eqnarray}
is figured in the first line  of Fig.\ref{Fig: BS}.

The coupled channel equation can be rewritten as two equations shown
in the second line. And we can construct the vertex with definite
quantum numbers as Eq.~(\ref{Eq:positive charged sytstem}).  After
flavor factors are isolated, two components of
$|{Z_{P\bar{P}^*}^{(T)}}^+\rangle$ should be the same under isospin
$SU(2)$ symmetry. Hence, after rearranging  the notations of the
momenta in two equations in the second line of Fig.\ref{Fig: BS} and
summing them up, we reach one equation at the last line of of
Fig.\ref{Fig: BS}. Here the flavor factors are isolated out as $I^i_d$
and $I^j_c$ for direct diagram and cross diagram with $i$ for
different exchanged light meson, which is the same as the flavor
factors used in the nonrelativistic OBE
model~\cite{Shen:2010ky,Hu2011}. In the existing OBE model
calculation, the final momenta of cross diagram are not exchanged
correctly~\cite{Tornqvist:2004qy,Sun:2011uh,Shen:2010ky}. This
exchange does not affect the nucleon-nucleon interaction in the study
of nuclear force, the $u/d$ quark-quark interaction in the consistent
quark model and $P\bar{P}/P^*\bar{P}^*$ system due to the same masses
and spins of nucleon ,$u/d$ quark and the $P\bar{P}/P^*\bar{P}^*$
under $SU(2)$ symmetry. However, for the $P\bar{P}^*$ system composed of
two constituents with different masses and spins, it is essential to
make such exchange in the cross diagram.

\subsection{Quasipotential approximation in covariant spectator theory}

The BSE of the vertex function $\Gamma$
for a system composed of a vector meson and a
pseudoscalar meson (marked as constituent 1 and 2) is written explicitly as
\begin{eqnarray}
	&&|\Gamma^\mu\rangle={\cal V}^{\mu\nu}~ G_{\nu\mu'}~ |\Gamma^{\mu'}\rangle,
\end{eqnarray}
where  the propagator is
\begin{eqnarray}
G^{\mu'\mu}=G_1^{\mu'\mu}~G_2
=\frac{-P_1^{\mu'\mu}}{(k_1^2-m_1^2)(k_2^2-m_2^2)}\equiv P_1^{\mu'\mu}
G_0,
\end{eqnarray}
where $P_1^{\mu\nu}=-g^{\mu\nu}+\frac{k_1^\mu k_1^\nu}{m_1^2}$ and $k_{1,2}$ and $m_{1,2}$ are the momentum and mass for constituent 1 or 2.

As in the study of nucleon-nucleon interaction, a quasipotential
approach should be used to reduce 4-dimension equation to 3-dimension
equation. Here we adopt the covariant spectator theory which is suitable to study a system
with different constituents~\cite{Gross:2010qm}. The
heavy constituent, here the vector meson marked as constituent 1, is
treated as
on-shell, so the numerator of propagator can be written as
$P_1^{\mu\nu}=\epsilon_{1\lambda}^\mu\epsilon_{1\lambda}^\nu$ with $\lambda$ being
the helicity of vector meson. Here and hereafter the sum notation about helicity $\lambda$ is omitted.
After multiplying the polarized vector
$\epsilon^\mu_{1\lambda}$, we have
\begin{eqnarray}
	&&|{\Gamma}_{\lambda}\rangle
	={\cal{V}}_{\lambda\lambda'}~{G}_0~
	|{\Gamma}_{\lambda'}\rangle,
\end{eqnarray}
with
$|{\Gamma}_{\lambda}\rangle=\epsilon^\mu_{1\lambda}|{\Gamma}_\mu\rangle$
and ${\cal{V}}_{\lambda\lambda'}=\epsilon^\mu_{1\lambda}\cdot{{\cal V}_{\mu\nu}}\cdot
\epsilon^{\nu}_{1\lambda'}$.

The propagator written down in
the center of mass frame where $p=(W,{\bm 0})$  is
\begin{eqnarray}
G_0=2\pi i\frac{\delta^+(k_1^2-m_1^2)}{k_2^2-m_2^2}=2\pi
i\frac{\delta^+(k^0_1-E_1))}{2E_1[(W-E_1)^2-E_2^2]},
\end{eqnarray}
where  $E_{1,2}=\sqrt{m_{1,2}^2+|\bm k|^2}$.
After moving a factor
\begin{eqnarray}
A=\sqrt{\frac{2E_2}{W-E_1+E_2}}
\end{eqnarray}
to potential kernel
and vertex, we have ${\cal	\bar{V}}_{\lambda\lambda'}=A{\cal{V}}_{\lambda\lambda'}A'$,
$|\bar{\Gamma}_{\lambda}\rangle=A|\Gamma_{\lambda}\rangle$
and $\bar{G}_0=G_0/A^2$.
The vertex function can be related to the Bethe-Salpeter
bound state wave function as
$|\psi_{\lambda}\rangle=\bar{G}_0|\bar{\Gamma}_{\lambda}\rangle$ and
we reach the BSE for the wave function
\begin{eqnarray}
\bar{G}^{-1}_0~|\psi_{\lambda}\rangle=\bar{\cal
V}_{\lambda\lambda'}|\psi_{\lambda'}\rangle.
\end{eqnarray}

The normalization of the wave function can be obtained by the
normalization of the vertex,
\begin{eqnarray}
	1&=&i\frac{\langle\Gamma^\mu|G^{\mu\nu}-G^{\mu\mu'} {\cal V}^{\mu'\nu'} G^{\nu'\nu}|\Gamma^\nu\rangle}{p^2-M^2}\nonumber\\
	&=&i\langle
	\psi_{\lambda}|(-iN^2\delta_{\lambda\lambda'}-\bar{\cal V}_{\lambda\lambda'})'|\psi_{\lambda'}\rangle.
\end{eqnarray}
Here, $\psi\to0$ when $|{\bm k}|\to\infty$ and $A$ is stable with small $|{\bm k}|$. As usual we assume
the dependence of ${\cal V}$ on $W$ is small. Hence, $(\bar{\cal V})'$
is negligible.
The normalized wave functions can be introduced
as $|\phi\rangle=N|{\psi}\rangle$ with $N=\sqrt{2E_12E_2}/\sqrt{(2\pi)^52W}$.
The integral equation  can be written explicitly as
\begin{eqnarray}
	(W-E_1({\bm k})-E_2({\bm k}))\phi_{\lambda}({\bm k})=\int\frac{d{\bm k}'}{(2\pi)^3}V_{\lambda\lambda'}({\bm k},{\bm k}',W)\phi_{\lambda'}({\bm k}'),\nonumber\\
\end{eqnarray}
with
\begin{eqnarray}
V_{\lambda\lambda'}({\bm k},{\bm k}',W)=\frac{i~\bar{\cal
	V}_{\lambda\lambda'}({\bm k},{\bm k}',W)}{\sqrt{2E_1({\bm k})2E_2({\bm k})2E'_1({\bm k}')2E'_2({\bm k}')}}.
\end{eqnarray}

The BSE can be related to the nonrelativistic OBE model
~\cite{Tornqvist:2004qy,Sun:2011uh} by nonrelativization and the Fourier
transformation.  The reduced equation is
\begin{eqnarray}
[\frac{\nabla^2}{2\mu}-E]\phi({\bm r})=\int d{\bm r} V({\bm r},{\bm r}')\phi({\bm r}'),\label{Sch}
\end{eqnarray}
where $\mu$ is reduced mass and $E=m_1+m_2-W$ is the binding energy. The potential in coordinate space can be defined as
\begin{eqnarray}
V({\bm r},{\bm r}')=\frac{1}{(2\pi)^3}\frac{1}{2}\int d{\bm q} d{\bm
q}' e^{i[{\bm q}'\cdot\frac{{\bm r}-{\bm r}'}{2}-{\bm
q}\cdot\frac{{\bm r}+{\bm r}'}{2}]}V({\bm k},{\bm k}'),
\end{eqnarray}
where ${\bm q}={\bm k}'-{\bm k}$ and ${\bm q}'={\bm k}'+{\bm k}$.  For
direct diagram the potential after nonrelativization can be
written as $V({\bm q})$ which can be transformed to $V(\bm
r)\delta({\bm r}-{\bm r'})$ in coordinate space. The Schr\"odinger
equation can be obtained. For cross diagram the potential is in
the form $V({\bm q}')$, which is transformed to $V(\bm r)\delta({\bm
r}+{\bm r'})$, so the wave functions $\phi$ in the two sides of
Eq.~(\ref{Sch}) are about ${\bm r}$ and $-{\bm r}$. It is no longer feasible
to treat this issue with the Shr\"odinger equation. However, the same
treatment for direct diagram and cross diagram was mistakenly applied
by many authors ~\cite{Tornqvist:1991ks,Tornqvist:2004qy,Sun:2011uh}.
In the current paper, we do not make such nonrelativization but partial wave expansion of BSE.

\subsection{Partial wave expansion}
The wave function has an angular dependence as
\begin{eqnarray}
&&\phi_{\lambda}({\bm
k})=\sqrt{\frac{2J+1}{4\pi}}D^{J*}_{\lambda_R,\lambda}(\phi,\theta,0)\phi^J_{\lambda,\lambda_R}(|{\bm
k}|),
\end{eqnarray}
where $J$ partial wave is considered and $D^{J*}_{\lambda_R,\lambda}(\phi,\theta,0)$ is the rotation matrix with $\lambda_R$ being the helicity of bound state.

The potential after partial wave expansion is~\cite{Gross:2008ps}
\begin{eqnarray}
V_{\lambda\lambda'}^J(|{\bm k}|,|{\bm k}|')=2\pi\int
d\cos\theta_{k,k'}
d^{J}_{\lambda,\lambda'}(\theta_{k,k'})V_{\lambda\lambda'}({\bm
k},{\bm k}'),
\end{eqnarray}
where $\theta_{k,k'}$ is angle between ${\bm k}$ and ${\bm k}'$.

Now we reach a one-dimensional integral equation,
\begin{eqnarray}
	&&(W-E_1(|\bm k|)-E_2(|\bm k|))\phi^J_{\lambda}(|{\bm
	k}|)\nonumber\\&=&\int  \frac{|{\bm k}'|^2d|{\bm
	k}'|}{(2\pi)^3}V^J_{\lambda\lambda'}(|{\bm k}'|,|{\bm
	k}'|)\phi^J_{\lambda'}(|{\bm k}'|).\label{Eq: final equation}
\end{eqnarray}

For the wave function, we have relations
$\phi_\lambda=\phi_{-\lambda}$ for system with quantum number
$J^P=1^+,0^-$ and $\phi_\lambda=-\phi_{-\lambda}$ for system with
$1^-$. For potential we have
$V_{\lambda\lambda'}=V_{-\lambda-\lambda'}$. Hence, we can only
consider the independent wave functions and potentials. In this paper
we choose $\phi_1=\sqrt{2}\phi_\pm$ and $\phi_0$ for system with
$J^P=1^+$, $\phi_0$ for system with $J^P=0^-$ and
$\phi_1=\sqrt{2}\phi_\pm$ for system with $J^P=1^-$ with proper
normalization.

Hence, we have the coupled equation,
\begin{eqnarray}
	&&(W-E_1({\bm k})-E_2({\bm k}))\phi^J_{i}(|{\bm k}|)\\\nonumber&=&\sum_{j}\int  \frac{|{\bm k}'|^2d|{\bm k}'|}{(2\pi)^3}V^J_{ij}(|{\bm k}'|,|{\bm k}'|)\phi^J_{j}(|{\bm k}'|),
\end{eqnarray}
with the normalization as
\begin{eqnarray}
\sum_{i}\int |{\bm k}|^2d|{\bm k}|\phi^{J~2}_{i}(|{\bm k}|)=1.
\end{eqnarray}

 The potential $V_{ij}$ can be written with $V^J_{\lambda\lambda'}$ as
\begin{eqnarray}
V(1^+)&=&\left(\begin{array}{cc} V^1_{11}+V^1_{1-1}  & \sqrt{2}V^1_{10} \\ \frac{1}{\sqrt{2}} (V^1_{01}+V^1_{0-1} )   &  V^1_{00} \end{array}\right),\nonumber\\
V(1^-)&=&V^1_{11}-V^1_{1-1},\nonumber\\
V(0^-)&=&V^0_{00}.
\end{eqnarray}

\subsection{Numerical solution of the BSE}

The coupled one-dimensional integral Eq.~(\ref{Eq: final equation})
can be rewritten in the form as
\begin{eqnarray}
	&&\phi^J_{i}(|{\bm	k}|)=\int  d|{\bm k}'|A_{ij}(|{\bm k}'|,|{\bm
	k}'|)\phi^J_{j}(|{\bm k}'|).\label{Eq: final equation}
\end{eqnarray}

To solve the integral equation, we discrete the $|{\bm k}|$ and $|{\bm
k}'|$ to $|{\bm k}|_k$ and $|{\bm k}|_{k'}$ by the Gauss quadrature , then
the above equation can be transformed to a matrix equation,
\begin{eqnarray}
	W\phi_{ik}=\sum_{jk'}A_{ik,jk'}(W)\omega_j\phi_{jk'}.
\end{eqnarray}
We remark the indices $ik$ and $jk'$ to new indices $i$ and $j$ and absorb $\omega$ to $A$,and have
\begin{eqnarray}
	W\phi_{i}=\sum_{j}\tilde{A}_{ij}(W)\phi_j,
\end{eqnarray}
which can be written in a compact matrix form,
\begin{eqnarray}
	W\phi=\tilde{A}(W)\phi.\label{Eq: matrix equation}
\end{eqnarray}

Due to the dependence of the total energy $W$ of $\tilde{A}$, the
solution of above equation~(\ref{Eq: matrix equation}) is a nonlinear
spectral problem.  In this paper  the recursion method in
Ref~\cite{Soloveva:2001aa,He:2013oma} is adopted.  The values of $W$
and $\psi$ are obtained by performing a sequence of approximations by
using the recursion relation,
\begin{eqnarray}
	W^{(l)}_n\psi=\tilde{A}(W^{(l-1)}_s)\psi,
\end{eqnarray}
where the upper index $(l)$ and the lower index $n=1,2,\cdots
s,\cdots$ are the iteration number and  eigenvalue number, respectively. At
the first iteration, we choose a sought eigenvalue and  substitute it
into the kernel $\tilde{A}$.  In the current problem, the binding
energy is very small, so the total energy $W$ is chosen as the sum of
the masses of two constituents $m_1+m_2$. The $n{\rm th}$ eigenvalues
can be obtained with the help of the code of DGEEV function in the NAG
Fortran Library.  If we want to obtain the $s{\rm th}$ eigenvalues,
the corresponding eigenvalue should be substituted in kernel
$\tilde{A}$ on each iterative loop. In the current paper, we choose ground
state.  With new kernel, the linear spectral problem is solved again.
On each iteration stopping criterion
$|W^{(l)}_s-W_s^{(l-1)}|<\epsilon$, which is related to precision of
the final results, is tested.  Once stopping criterion fulfilled, the
iterative process is terminated. The eigenvalue $W_s^{(l)}$ and
eigenfunction $\phi_s^{(l)}$ obtained on the last iteration are
adopted as the solution.

\section{ The one-boson-exchange potential}

To give the potential, we adopt the effective Lagrangians of
the pseudoscalar and vector mesons with heavy flavor mesons from the
heavy quark effective theory
~\cite{Colangelo:2003sa,Casalbuoni:1996pg},
\begin{eqnarray*}\label{eq:lag-p-exch}
  \mathcal{L}_{P^*P\mathbb{P}} &=&
  -\frac{2g\sqrt{m_Pm_{P^*}}}{f_\pi} (P_bP^{*\dag}_{a\lambda}+P^*_{b\lambda}P^\dag_{a})\partial^\lambda{}\mathbb{P}_{ba}
  \nonumber\\&+&\frac{2g\sqrt{m_Pm_{P^*}}}{f_\pi}
  (\tilde{P}^{*\dag}_{a\lambda}\tilde{P}_b+\tilde{P}^\dag_{a}\tilde{P}^*_{b\lambda})\partial^\lambda\mathbb{P}_{ab},\\
  \mathcal{L}_{P^*P\mathbb{V}} &=&
  -i\sqrt{2}\lambda g_V\varepsilon_{\lambda\alpha\beta\mu}
  (P^{*\mu\dag}_a\overleftrightarrow{\partial}^\lambda P_b
  +P^\dag_a\overleftrightarrow{\partial}^\lambda P_b^{*\mu})
  (\partial^\alpha{}\mathbb{V}^\beta)_{ba}\nonumber\\
  &-&i\sqrt{2}\lambda g_V\varepsilon_{\lambda\alpha\beta\mu}
  ( \tilde{P}^{*\mu\dag}_a\overleftrightarrow{\partial}^\lambda
  \tilde{P}_b  +\tilde{P}^\dag_a\overleftrightarrow{\partial}^\lambda
  \tilde{P}_b^{*\mu})(\partial^\alpha{}\mathbb{V}^\beta)_{ab},\nonumber\\
	\mathcal{L}_{PP\mathbb{V}} &=& -i\frac{\beta
	g_V}{\sqrt{2}} P_a^\dag\overleftrightarrow{\partial}_\mu P_b\mathbb{V}^\mu_{ba}
	+i\frac{\beta	g_V}{\sqrt{2}}\tilde{P}_a^\dag
	\overleftrightarrow{\partial}_\mu \tilde{P}_b\mathbb{V}^\mu_{ab}, \nonumber\\
  \mathcal{L}_{P^*P^*\mathbb{V}} &=& i\frac{\beta g_V}{\sqrt{2}}
  P_a^{*\dag}\overleftrightarrow{\partial}_\mu P^*_b\mathbb{V}^\mu_{ba}
  \nonumber\\&-&i2\sqrt{2}\lambda
  g_V m_{P^*}P^{*\mu}_bP^{*\nu\dag}_a(\partial_\mu\mathbb{V}_\nu-\partial_\nu\mathbb{V}_\mu)_{ba}
  \nonumber\\
  &-& i\frac{\beta
  g_V}{\sqrt{2}}\tilde{P}_a^{*\dag}\overleftrightarrow{\partial}_\mu
  \tilde{P}^*_b\mathbb{V}^\mu_{ab}
  \nonumber\\&-&i2\sqrt{2}\lambda  g_Vm_{P^*}\tilde{P}^{*\mu\dag}_a\tilde{P}^{*\nu}_b(\partial_\mu\mathbb{V}_\nu-\partial_\nu\mathbb{V}_\mu)_{ab}
,\\
  \mathcal{L}_{PP\sigma} &=&
  -2g_\sigma m_{P}P_a^\dag P_a\sigma -2g_\sigma m_{P}\tilde{P}_a^\dag \tilde{P}_a\sigma, \nonumber\\
  \mathcal{L}_{P^*P^*\sigma} &=&
  2g_\sigma m_{P^*}P_a^{*\dag} P^*_a\sigma +2g_\sigma m_{P^*}\tilde{P}_a^{*\dag}
  \tilde{P}^*_a\sigma.
\end{eqnarray*}
where the octet pseudoscalar and nonet vector
meson matrices read as
\begin{eqnarray}
\mathbb{P}&=&\left(\begin{array}{ccc}
\frac{\pi^{0}}{\sqrt{2}}+\frac{\eta}{\sqrt{6}}&\pi^{+}&K^{+}\\
\pi^{-}&-\frac{\pi^{0}}{\sqrt{2}}+\frac{\eta}{\sqrt{6}}&
K^{0}\\
K^- &\bar{K}^{0}&-\frac{2\eta}{\sqrt{6}}
\end{array}\right),\nonumber\\
\mathbb{V}&=&\left(\begin{array}{ccc}
\frac{\rho^{0}}{\sqrt{2}}+\frac{\omega}{\sqrt{2}}&\rho^{+}&K^{*+}\\
\rho^{-}&-\frac{\rho^{0}}{\sqrt{2}}+\frac{\omega}{\sqrt{2}}&
K^{*0}\\
K^{*-} &\bar{K}^{*0}&\phi
\end{array}\right).\label{vector}
\end{eqnarray}
Here we choose parameters $g=0.59$, $\beta$=0.9, $\lambda$=0.56 GeV$^{-1}$, $g_V=5.8$ and $g_\sigma = g_\pi/(2\sqrt{6})$ with $g_\pi = 3.73$~\cite{Isola:2003fh, Falk:1992cx}.

With above Lagrangians, the potential kernel ${\cal V}$ can be written as
\begin{eqnarray}
{\cal V}^{Direct}_{\mathbb{V}}&=&i\frac{\beta^2g^2_V}{2}[
	\frac{(k_1+k'_1)\cdot(k_2+k'_2)}{q^2-m_\mathbb{V}^2}]\epsilon_1\cdot\epsilon'_1\nonumber\\
	{\cal V}^{Direct}_{\sigma}&=&i4g^2_\sigma m_{P}m_{P^*}\frac{\epsilon_1\cdot\epsilon'_1}{q^2-m^2_\sigma}\nonumber\\
	{\cal V}^{Cross}_{\mathbb{P}}n
	&=&i\frac{4g^2m_Pm_{P^*}}{f^2_\pi}
	\frac{q\cdot\epsilon_1 q\cdot\epsilon'_1}{q^2-m^2_\mathbb{P}}\nonumber\\
{\cal V}^{Cross}_\mathbb{V}
&=&i8\lambda^2
g^2_V \frac{1}{q^2-m_\mathbb{V}^2}(q\cdot\epsilon_1 q\cdot\epsilon'_1 k_2\cdot k'_2\nonumber\\
&+&\epsilon_1\cdot\epsilon'_1
(k_2\cdot q k'_2\cdot q-k_2\cdot k'_2 q^2))
	\end{eqnarray}
where $q=k'_1-k_1$ for direct diagram and $q=k'_2-k_1=k_2-k'_1$ for
cross diagram.  A form factor
$h(k^2)=\frac{\Lambda^4}{(m^2-k^2)^2+\Lambda^4}$ is introduced to
compensate for the off-shell effect of heavy meson.  In the propagator of
exchanged meson we make a replacement $q^2\to-|q^2|$ to remove
singularities as Ref.~\cite{Gross:2008ps}. The form factor for
light meson is chosen as
$f(q^2)=\frac{\Lambda^2-m^2}{\Lambda^2+|q^2|}$. The flavor factors
$I^i_d$ and $I^i_c$ for direct and cross diagrams are presented in the
following table \ref{flavor factor}.

\renewcommand\tabcolsep{0.32cm}
\renewcommand{\arraystretch}{1.4}
\begin{table}[hbtp!]
\caption{The flavor factors $I^i_d$ and $I^i_c$ for direct and cross diagrams and different exchange mesons.
\label{flavor factor}.}
\begin{center}
	\begin{tabular}{cccccccc}\bottomrule[2pt]
&\multicolumn{3}{c}{Direct diagram} &\multicolumn{4}{c}{Cross diagram}\\
&\multicolumn{3}{c}{\scalebox{0.15}{\includegraphics[ bb=100 500 500 750,clip]{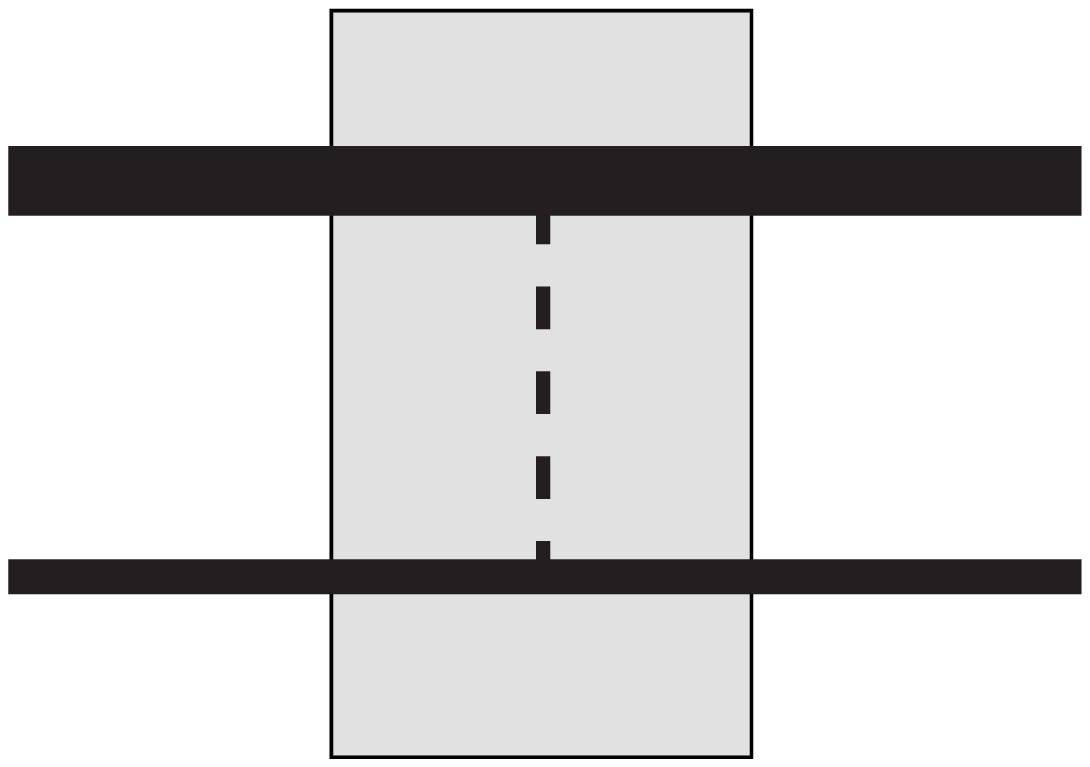}}}
	&\multicolumn{4}{c}{\scalebox{0.15}{\includegraphics[ bb=80 500 520 750,clip]{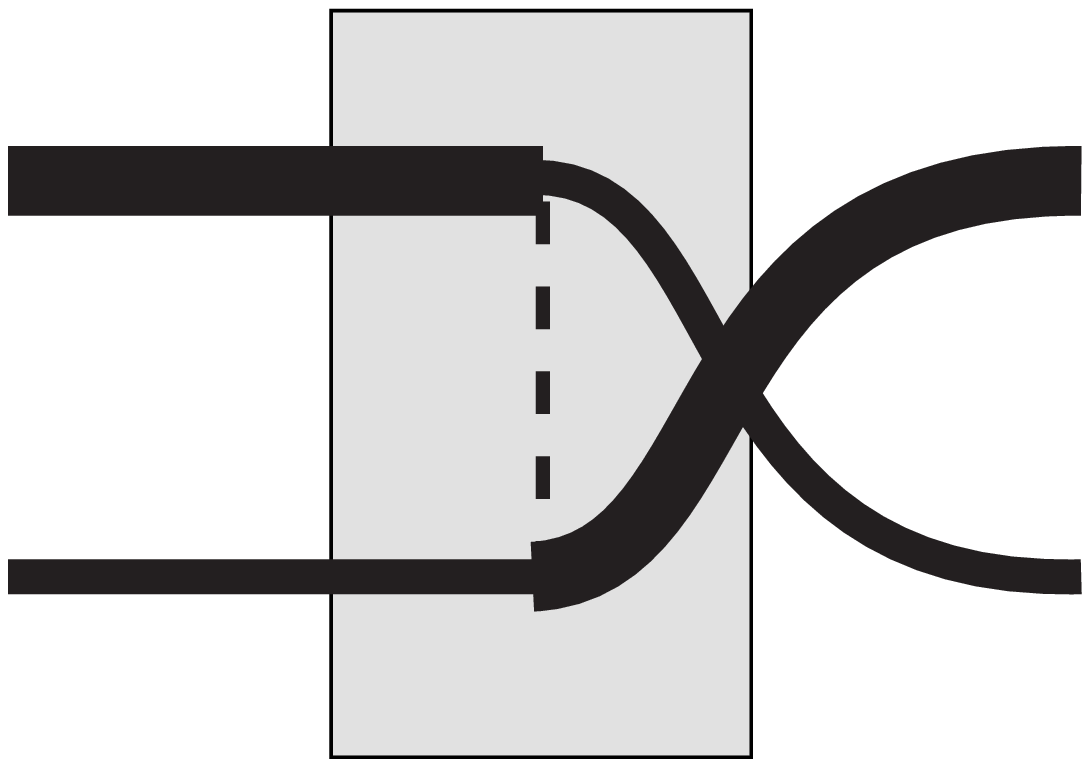}}}
\\\cline{2-4}\cline{5-8}
Exchanged&\multicolumn{2}{c}{$\mathbb{V}$}&$\mathbb{S}$
&\multicolumn{2}{c}{$\mathbb{P}$}&\multicolumn{2}{c}{$\mathbb{V}$}\\\cline{2-8}
meson  &$\rho$ &$\omega$
&$\sigma$&$\pi$
&\multicolumn{1}{c}{$\eta$}  &$\rho$
&\multicolumn{1}{c}{$\omega$} \\\hline
${Z_{P\bar{P}^*}^{(T)}}$
&$-\frac{1}{2}$&$\frac{1}{2}$&1 &$-\half$
&\multicolumn{1}{c}{$\frac{1}{6}$}
&$-\half$  &\multicolumn{1}{c}{$\half$} \\
${Z_{P\bar{P}^*}^{(S)}}$
&$\thalf$&$\half$& $1$&$-\frac{3}{2}$ &
\multicolumn{1}{c}{$\frac{1}{6}$} & $\thalf$
&\multicolumn{1}{c}{$\frac{1}{2}$} \\
\toprule[2pt]
\end{tabular}
\end{center}

\end{table}

\section{Numerical results}

In this paper, the states with $J=0,1$
are considered and the results of the binding energy $E=m_1+m_2-W$ are listed in Table~\ref{diagrams}
with cutoffs in the range $0.8<\Lambda<5$ GeV and compared with the nonrelativistic OBE model~\cite{Sun:2011uh,Sun:2012zzd}.
In this paper, only bound states with small binding energy $E<10$
MeV are considered because hadronic molecular state is defined as a loosely
bound state.
\renewcommand\tabcolsep{0.225cm}
\renewcommand{\arraystretch}{0.7}
\begin{table}[hbtp!]
\caption{The binding energies $E$ for $D\bar{D}^*$ and $B\bar{B}^*$
systems with different cutoff $\Lambda$
	obtained in this work (BS) and in the nonrelativistic OBE model (OBE)~\cite{Sun:2011uh} .
	``--'' means that no bound state is found. The cutoff
	$\Lambda$ and binding energy are in the units of GeV and
	MeV, respectively.
\label{diagrams}.}
\begin{center}
	\begin{tabular}{crrrrrrrrccccc}\bottomrule[2pt]
		&  \multicolumn{4}{c}{$D\bar{D}^*$} &
  \multicolumn{4}{c}{$B\bar{B}^*$} \\\hline
  &  \multicolumn{2}{c}{BS} &  \multicolumn{2}{c}{OBE} &
  \multicolumn{2}{c}{BS} &  \multicolumn{2}{c}{OBE}\\\hline
$I^G(J^{PC})$   &  $\Lambda$ & $E$ &  $\Lambda$ & $E$ &  $\Lambda$ &
$E$ &  $\Lambda$ & $E$ \\\bottomrule[1pt]
$0^-(0^{--})$&  -- & -- && & 1.5  & 1.6 \\
&  -- & -- && & 1.7  & 4.1 \\
&  -- & -- && & 1.9  & 6.7 \\
$0^+(0^{-+})$& --  & -- &&& --&-- \\
$0^-(1^{--})$& --  & -- &&& 1.6 & 1.4\\
& --  & -- &&& 1.7 & 3.7\\
& --  & -- &&& 1.8 & 6.4\\
$0^+(1^{-+})$& --  &--  &&& --& -- \\
$0^-(1^{+-})$& 1.3  & 0.2 & 1.4 & 3.4 & 1.1 & 0.6 &1.4&1.6 \\
& 1.4  & 6.0 & 1.5 &16.6 & 1.2 & 7.8&1.5 &13.0 \\
$0^+(1^{++})$& 2.0  & 0.2 &1.1 &0.6 & 1.3  &0.2 &1.1 &0.6\\
& 2.2  & 1.4 &1.2& 4.4 & 1.5  &3.0 &1.2 &4.4\\
& 2.4  & 4.1 &1.3 &11.8 & 1.7  &7.4 &1.3&11.8\\\hline
$1^+(0^{-})$& --  &--  &&& --& --\\
$1^-(0^{-})$& --  &--  &&& --& --\\
$1^+(1^{-})$& --  &-- && & --& --\\
$1^-(1^{-})$& --  &--  &&& --& --\\
$1^+(1^{+})$& --  &-- &--&-- & --& --&2.1&0.2\\
& --  &-- &--&-- & --& --&2.2&1.6\\
& --  &-- &--&-- & --& --&2.5&4.7\\
$1^-(1^{+})$& --  &-- &--&-- & --& -- &4.9 &0.1\\
& --  &-- &--&-- & --& -- &5.0 &0.4\\
\toprule[2pt]
\end{tabular}
\end{center}

\end{table}

In the isoscalar vector, there exist the hidden bottomed bound state
solutions with quantum number $I^G(J^{PC})=0^-(0^{--})$,
$0^-(1^{--})$, $0^-(1^{+-})$ and $0^-(1^{++})$  and  the hidden
charmed bound state solution with $0^-(1^{+-})$ and $0^-(1^{++})$.
The $D\bar{D}^*$ bound state with $I^G(J^{PC})=0^+(1^{++})$ can be
related to the $X(3872)$. This bound state was also found in the
nonrelativistic OBE model~\cite{Sun:2011uh} but with different
cutoffs. It is well known that there exists a $c\bar{c}$ component in
X(3872) ~\cite{Coito:2012vf}. In the current paper only the hadronic
molecular states are considered. The discussion about the $c\bar{c}$
component is beyond the scope of this work and not considered
here.

In the isovector sector, the nonrelativistic OBE model predicted a
molecular state $B\bar{B}^*$ with $I^G(J^P)=1^+(1^{+})$ with cutoffs
about 2 GeV, which is assigned to the observed $Z_b(10610)$
state~\cite{Sun:2011uh}. The observed $Z_c(3900)$ is also explained as
$D\bar{D}^*$ state in literatures~\cite{Wang:2013cya,Guo:2013sya}. In
our calculation, there does not exist isovector bound state solution
with all cutoffs $\Lambda<5$ GeV, which suggests the structures
$Z_b(10610)$ and  $Z_c(3900)$ should be originated from other
mechanisms, such as four-quark states or cusp
effect~\cite{Chen:2011pv}. In fact, if we compared the experimental
masses for $Z_b(10610)$ and $Z_c(3900)$ and the thresholds for
$B\bar{B}^*$ and $D\bar{D}^*$, we can find that these two states are
above the thresholds, which conflicts with the molecular state
assumption. The lattice results disfavored the possibility of a
shallow bound state for $D\bar{D}^*$ interaction
also~\cite{Prelovsek:2013xba,Chen:2014afa}.

\section{Summary}

In this paper the $B\bar{B}^*$ and $D\bar{D}^*$ systems are studied in
a BSE approach with quasipotential potential approximation by adopting
the covariant spectator theory which is suitable to study a system
with different constituents. In our calculation, both  direct and
cross diagrams are considered in the one-boson-exchange potential so
that the $\pi$ exchange which was found more important in the
$B\bar{B}^*$ and $D\bar{D}^*$
interactions~\cite{Sun:2011uh,Sun:2012zzd} is included. Partial wave
expansion is used to reduce the BSE to a one-dimensional equation, which
is solved by a  recursion method. The numerical results indicate the
existence of an isoscalar bound state $D\bar{D}^*$ with
$J^{PC}=1^{++}$, which may be related to the $X(3872)$. In the
isovector sector, no bound state is produced from both $D\bar{D}^*$
and $B\bar{B}^*$ interactions, which disfavors the molecular state
explanations for $Z_b(10610)$ and $Z_c(3900)$.

It is found in our calculation that for cross diagram  the BSE can not be transformed to the
schr\"odinger equation with potential in coordinate space $V({\bm
r})$. This problem appears in all systems composed of two constituents
with different masses and /or spins which can convert to each other,
such as the $K$ exchange potential between $s$ quark and $u/d$ quark
in the constituent quark model~\cite{Glozman:1995fu} and $NN^*$
interaction~\cite{Zhao:2013xha} where a potential in coordinate space
are used.  The results obtained in this work show there does not exist
isovector $B\bar{B}^*/D\bar{D}^*$ bound state, which is more
consistent with the experiment and the lattice
QCD~\cite{Prelovsek:2013xba,Chen:2014afa}.  Hence, one should be
cautious in the direct application of potential in coordinate space
obtained by a simple Fourier transformation, which has been widely
used in the studies of the hadron spectrum, hadronic molecular states
and other fields~\cite{Glozman:1995fu,Tornqvist:1991ks,Sun:2011uh}.

\acknowledgements
This project is partially supported by the Major State
Basic Research Development Program in China (No. 2014CB845405),
the National Natural Science
Foundation of China (Grants No. 11275235, No. 11035006)
and the Chinese Academy of Sciences (the Knowledge Innovation
Project under Grant No. KJCX2-EW-N01).


\begin{thebibliography}{31}



  	
		
\expandafter\ifx\csname natexlab\endcsname\relax\def\natexlab#1{#1}\fi
\expandafter\ifx\csname bibnamefont\endcsname\relax
  \def\bibnamefont#1{#1}\fi
\expandafter\ifx\csname bibfnamefont\endcsname\relax
  \def\bibfnamefont#1{#1}\fi
\expandafter\ifx\csname citenamefont\endcsname\relax
  \def\citenamefont#1{#1}\fi
\expandafter\ifx\csname url\endcsname\relax
  \def\url#1{\texttt{#1}}\fi
\expandafter\ifx\csname urlprefix\endcsname\relax\def\urlprefix{URL }\fi
\providecommand{\bibinfo}[2]{#2}
\providecommand{\eprint}[2][]{\url{#2}}

\bibitem{Tornqvist:1991ks}
  N.~A.~Tornqvist,
  Phys.\ Rev.\ Lett.\  {\bf 67}, 556 (1991).
	
\bibitem{Choi:2003ue}
  S.~K.~Choi {\it et al.}  [Belle Collaboration],
  Phys.\ Rev.\ Lett.\  {\bf 91}, 262001 (2003)
  [hep-ex/0309032].

\bibitem{Tornqvist:2004qy}
  N.~A.~Tornqvist,
  Phys.\ Lett.\ B {\bf 590}, 209 (2004)
  [hep-ph/0402237].

\bibitem{Close:2003sg}
  F.~E.~Close and P.~R.~Page,
  Phys.\ Lett.\ B {\bf 578}, 119 (2004)
  [hep-ph/0309253].

\bibitem{AlFiky:2005jd}
  M.~T.~AlFiky, F.~Gabbiani and A.~A.~Petrov,
  Phys.\ Lett.\ B {\bf 640}, 238 (2006)
  [hep-ph/0506141].

\bibitem{Belle:2011aa}
  A.~Bondar {\it et al.}  [Belle Collaboration],
  Phys.\ Rev.\ Lett.\  {\bf 108}, 122001 (2012)
  [arXiv:1110.2251 [hep-ex]].

\bibitem{Ablikim:2013mio}
  M.~Ablikim {\it et al.}  [BESIII Collaboration],
  Phys.\ Rev.\ Lett.\  {\bf 110}, 252001 (2013)
  [arXiv:1303.5949 [hep-ex]].

\bibitem{Sun:2011uh}
  Z.~F.~Sun, J.~He, X.~Liu, Z.~G.~Luo and S.~L.~Zhu,
  Phys.\ Rev.\ D {\bf 84}, 054002 (2011)
  [arXiv:1106.2968 [hep-ph]].

\bibitem{Sun:2012zzd}
  Z.~F.~Sun, Z.~G.~Luo, J.~He, X.~Liu and S.~L.~Zhu,
  Chin.\ Phys.\ C {\bf 36}, 194 (2012).

\bibitem{Gross:2010qm}
  F.~Gross and A.~Stadler,
  Phys.\ Rev.\ C {\bf 82}, 034004 (2010)
  [arXiv:1007.0778 [nucl-th]].

\bibitem{Hummel:1989qn}
  E.~Hummel and J.~A.~Tjon,
  Phys.\ Rev.\ Lett.\  {\bf 63}, 1788 (1989).

\bibitem{Ke:2012gm}
  H.~W.~Ke, X.~Q.~Li, Y.~L.~Shi, G.~L.~Wang and X.~H.~Yuan,
  JHEP {\bf 1204}, 056 (2012)
  [arXiv:1202.2178 [hep-ph]].

\bibitem{He:2011ed}
  J.~He and X.~Liu,
  Eur.\ Phys.\ J.\ C {\bf 72}, 1986 (2012)
  [arXiv:1102.1127 [hep-ph]].

\bibitem{He:2012zd}
  J.~He, D.~Y.~Chen and X.~Liu,
  Eur.\ Phys.\ J.\ C {\bf 72}, 2121 (2012)
  [arXiv:1204.6390 [hep-ph]].

\bibitem{Adam:1997cx}
  J.~Adam, Jr., J.~W.~Van Orden and F.~Gross,
  Nucl.\ Phys.\ A {\bf 640}, 391 (1998)
  [nucl-th/9710009].

\bibitem{Shen:2010ky}
  L.~L.~Shen, X.~L.~Chen, Z.~G.~Luo, P.~Z.~Huang, S.~L.~Zhu, P.~F.~Yu and X.~Liu,
  Eur.\ Phys.\ J.\ C {\bf 70}, 183 (2010)
  [arXiv:1005.0994 [hep-ph]].

\bibitem{Hu:2010fg}
  B.~Hu, X.~L.~Chen, Z.~G.~Luo, P.~Z.~Huang, S.~L.~Zhu, P.~F.~Yu and X.~Liu,
  Chin.\ Phys.\ C {\bf 35}, 113 (2011)
  [arXiv:1004.4032 [hep-ph]].

\bibitem{Gross:2008ps}
  F.~Gross and A.~Stadler,
  Phys.\ Rev.\ C {\bf 78}, 014005 (2008)
  [arXiv:0802.1552 [nucl-th]].



\bibitem{Soloveva:2001aa}
T.~M. Soloveva
 Comput. Phys. Commun. {\bf 136}, 208 (2001)


\bibitem{He:2013oma}
  J.~He and P.~L.~L¨¹,
  Nucl.\ Phys.\ A {\bf 919}, 1 (2013)
  [arXiv:1309.6718 [hep-ph]].


\bibitem{Colangelo:2003sa}
  P.~Colangelo, F.~De Fazio and T.~N.~Pham,
  Phys.\ Rev.\ D {\bf 69}, 054023 (2004)
  [hep-ph/0310084].

\bibitem{Casalbuoni:1996pg}
  R.~Casalbuoni, A.~Deandrea, N.~Di Bartolomeo, R.~Gatto, F.~Feruglio and G.~Nardulli,
  Phys.\ Rept.\  {\bf 281}, 145 (1997)
  [hep-ph/9605342].

\bibitem{Isola:2003fh}
  C.~Isola, M.~Ladisa, G.~Nardulli and P.~Santorelli,
  Phys.\ Rev.\ D {\bf 68}, 114001 (2003)
  [hep-ph/0307367].

\bibitem{Falk:1992cx}
  A.~F.~Falk and M.~E.~Luke,
  Phys.\ Lett.\ B {\bf 292}, 119 (1992)
  [hep-ph/9206241].

\bibitem{Coito:2012vf}
  S.~Coito, G.~Rupp and E.~van Beveren,
  Eur.\ Phys.\ J.\ C {\bf 73}, 2351 (2013)
  [arXiv:1212.0648 [hep-ph]].

\bibitem{Wang:2013cya}
  Q.~Wang, C.~Hanhart and Q.~Zhao,
  Phys.\ Rev.\ Lett.\  {\bf 111}, no. 13, 132003 (2013)
  [arXiv:1303.6355 [hep-ph]].

\bibitem{Guo:2013sya}
  F.~K.~Guo, C.~Hidalgo-Duque, J.~Nieves and M.~P.~Valderrama,
  Phys.\ Rev.\ D {\bf 88}, 054007 (2013)
  [arXiv:1303.6608 [hep-ph]].

\bibitem{Chen:2011pv}
  D.~Y.~Chen and X.~Liu,
  Phys.\ Rev.\ D {\bf 84}, 094003 (2011)
  [arXiv:1106.3798 [hep-ph]].

\bibitem{Prelovsek:2013xba}
  S.~Prelovsek and L.~Leskovec,
  Phys.\ Lett.\ B {\bf 727}, 172 (2013)
  [arXiv:1308.2097 [hep-lat]].

\bibitem{Chen:2014afa}
  Y.~Chen, M.~Gong, Y.~H.~Lei, N.~Li, J.~Liang, C.~Liu, H.~Liu and J.~L.~Liu {\it et al.},
  Phys.\ Rev.\ D {\bf 89}, 094506 (2014)
  [arXiv:1403.1318 [hep-lat]].

\bibitem{Glozman:1995fu}
  L.~Y.~Glozman and D.~O.~Riska,
  Phys.\ Rept.\  {\bf 268}, 263 (1996)
  [hep-ph/9505422].

\bibitem{Zhao:2013xha}
  L.~Zhao, P.~N.~Shen, Y.~J.~Zhang and B.~S.~Zou,
  Eur.\ Phys.\ J.\ A {\bf 49}, 59 (2013)
  [arXiv:1302.4790 [hep-ph]].

\end{thebibliography}
\end{document}